\documentclass{jpp}

\usepackage{graphicx}
\usepackage{natbib}

\ifCUPmtlplainloaded \else
  \checkfont{eurm10}
  \iffontfound
    \IfFileExists{upmath.sty}
      {\typeout{^^JFound AMS Euler Roman fonts on the system,
                   using the 'upmath' package.^^J}%
       \usepackage{upmath}}
      {\typeout{^^JFound AMS Euler Roman fonts on the system, but you
                   dont seem to have the}%
       \typeout{'upmath' package installed. JPP.cls can take advantage
                 of these fonts, if you use 'upmath' package.^^J}%
      }
  \else
  \fi
\fi

\ifCUPmtlplainloaded \else
  \checkfont{msam10}
  \iffontfound
    \IfFileExists{amssymb.sty}
      {\typeout{^^JFound AMS Symbol fonts on the system, using the
                'amssymb' package.^^J}%
       \usepackage{amssymb}%
         \let\leq=\leqslant
         
      }{}
  \fi
\fi

\ifCUPmtlplainloaded \else
  \IfFileExists{amsbsy.sty}
    {\typeout{^^JFound the 'amsbsy' package on the system, using it.^^J}%
     \usepackage{amsbsy}}
    {}
\fi

\newsavebox{\astrutbox}
\sbox{\astrutbox}{\rule[-5pt]{0pt}{20pt}}

\def\av#1{\left\langle#1\right\rangle}

\newcommand\cmminusone{\,{\rm cm}^{-1}}

\renewcommand\d{{\rm d}}

\newcommand\erg{\,{\rm erg}}

\newcommand\fzero{{f_0}}
\newcommand\gammaintil{\tilde\gamma_{\rm in}}
\newcommand\gammaouttil{\tilde\gamma_{\rm out}}
\newcommand\grad{{\nabla}}

\renewcommand\i{{\rm i}}

\newcommand\kpc{{\rm \,kpc}}
\newcommand\keV{{\rm \,keV}}
\newcommand\yr{{\rm \,yr}}
\newcommand\K{{\rm \,K}}
\newcommand\kb{k_{\rm B}}

\newcommand\kMRI{k_{\rm MRI}}
\newcommand\kMTI{k_{\rm MTI}}
\newcommand\kz{k_z}
\newcommand\kR{k_R}

\newcommand\kv{{\bf k}}

\renewcommand\mp{m_{\rm p}}
\newcommand\Msun{{M}_{\odot}}
\newcommand\Mtherm{M_{\rm therm}}
\newcommand\Mtrue{M_{\rm true}}
\newcommand\Mpc{{\rm \,Mpc}}
\newcommand\ninst{n_{\rm inst}}
\newcommand\ncool{n_{\rm cool}}
\newcommand\nMRI{n_{\rm MRI}}

\newcommand\nezero{n_{e0}}

\newcommand\pzero{{p_0}}
\newcommand\rs{{r_{\rm s}}}
\newcommand\sminusone{\,{\rm s}^{-1}}

\newcommand\tcool{t_{\rm cool}}
\newcommand\tff{t_{\rm ff}}
\newcommand\tMRI{t_{\rm MRI}}
\newcommand\tinst{t_{\rm inst}}

\newcommand\Tzero{T_0}
\newcommand\vz{{v_z}}
\newcommand\vR{{v_R}}

\newcommand\vphi{v_{\phi}}

\newcommand\vzerophi{v_{0\phi}}

\newcommand\vv{{\bf v}}

\newcommand\G{{\rm G}}

\newcommand\rhozero{\rho_0}

\newcommand\omegaA{\omega_{\rm A}}
\newcommand\omegaAsq{\omega^2_{\rm A}}

\newcommand\omegad{\omega_{\rm d}}

\newcommand\omegaBV{\omega_{\rm BV}}
\newcommand\omegaBVsq{\omega^2_{\rm BV}}

\newcommand\omegarot{\omega_{\rm rot}}
\newcommand\omegarotsq{\omega^2_{\rm rot}}
\newcommand\omegacmag{\omega_{\rm c,mag}}

\newcommand\omegacphisq{\omega^2_{\rm c,\phi}}

\newcommand\omegacphi{\omega_{\rm c,\phi}}

\newcommand\omegath{\omega_{\rm th}}

\newcommand\kelvin{\,{\rm K}} \let\K=\kelvin

\newcommand\de{\partial}

\newcommand\Bv{{\bf B}}
\newcommand\Bvzero{\Bv_0}
\newcommand\Bzero{B_0}
\newcommand\Bzerophi{B_{0\phi}}
\newcommand\BzeroR{B_{0R}}
\newcommand\Bzeroz{B_{0z}}

\newcommand\Bphi{B_{\phi}}
\newcommand\BR{B_{R}}
\newcommand\Bz{B_{z}}
\newcommand\In{{\rm I}_n}

\newcommand\Qv{{\bf Q}}
\newcommand\Qvzero{{\bf Q}_0}
\newcommand\Rn{{\rm R}_n}

\renewcommand\div{\nabla \cdot}
\newcommand\rot{\nabla \times}
\newcommand\e{{\rm e}}

\renewcommand\H{\mathcal{H}}
\renewcommand\L{\mathcal{L}}

\newcommand\omegaca{\omega_{\rm c,a}}

\newcommand{\idl}{{\sc idl }}
\newcommand{\fzroots}{{\sc fz\_roots}}

\defcitealias{Bia13}{BEN13}
\defcitealias{Nip13}{NP13}
\defcitealias{Nip14}{NP14}

\title[MRI in cluster cool cores]{Magnetorotational instability in cool cores of galaxy clusters}

\author[C. Nipoti {\it et al.}]
{
{C\ls A\ls R\ls L\ls O\ns  N\ls I\ls P\ls O\ls T\ls I$^1$}\thanks{Email address for correspondence: carlo.nipoti@unibo.it},\ns
{L.\ns P\ls O\ls S\ls T\ls I$^1$},\ns
{S.\ns E\ls T\ls T\ls O\ls R\ls I$^{2,3}$}
 \and
{M.\ns B\ls I\ls A\ls N\ls C\ls O\ls N\ls I$^4$}
}

\affiliation{$^1$Dipartimento di Fisica e Astronomia, Universit\`a di Bologna, viale Berti-Pichat 6/2, I-40127 Bologna, Italy\\[\affilskip]
$^2$INAF, Osservatorio Astronomico di Bologna, via Ranzani 1, I-40127 Bologna, Italy\\[\affilskip]
$^3$INFN, Sezione di Bologna, viale Berti-Pichat 6/2, I-40127 Bologna, Italy\\[\affilskip]
$^4$Institute of Astro and Particle Physics, University of Innsbruck, A-6020 Innsbruck, Austria}

\pubyear{2015}
\volume{xxx}
\pagerange{xxx--xxx}
\date{in original form March 17 2015. Accepted May 29 2015}
\begin{document}

\maketitle

\begin{abstract}
Clusters of galaxies are embedded in halos of optically thin,
gravitationally stratified, weakly magnetized plasma at the system's
virial temperature.  Due to radiative cooling and anisotropic heat
conduction, such intracluster medium (ICM) is subject to local
instabilities, which are combinations of the thermal, magnetothermal
and heat-flux-driven buoyancy instabilities.  If the ICM rotates
significantly, its stability properties are substantially modified
and, in particular, also the magnetorotational instability (MRI) can
play an important role.  We study simple models of rotating cool-core
clusters and we demonstrate that the MRI can be the dominant
instability over significant portions of the clusters, with possible
implications for the dynamics and evolution of the cool cores.  Our
results give further motivation for measuring the rotation of the ICM
with future X-ray missions such as ASTRO-H and ATHENA.
\end{abstract}


\section{Introduction}\label{sec:intro}

Galaxy clusters are embedded in gaseous halos at the system's virial
temperature ($10^7-10^8\K$). Such intracluster medium (ICM) is a
dilute, weakly magnetized plasma, stratified in the cluster
gravitational potential.  The question of the local stability of the
ICM is a fundamental piece in the puzzle of the evolution of clusters
and in particular of cool-core clusters, in which the cooling time of
the plasma in the central regions is shorter than the cluster age.
The results of linear-stability analysis and magneto-hydrodynamics
(MHD) simulations indicate that, due to the joint effect of radiative
cooling and anisotropic heat conduction, the ICM is subject to several
local instabilities, which are combinations of the thermal instability
(TI; \citealt{Fie65}), the magnetothermal instability (MTI;
\citealt{Bal00}) and the heat-flux-driven buoyancy instability (HBI;
\citealt{Qua08}). Cosmological hydrodynamic simulations suggest that
the X-ray emitting halos of clusters, though mainly pressure
supported, might rotate significantly \citep{Fan09,Lau12}, with a
contribution of rotation especially important in the cool cores of
relaxed clusters \citep{Nag13}.  Observationally, the most direct
signature of ICM rotation would be the spatially resolved measure of
shifted X-ray emission line centroids. Unfortunately, given the
relatively poor spectral resolution ($\gtrsim 100$ eV) of currently
available X-ray instruments, so far there are no meaningful
constraints on rotation based on such a measure, which requires the
spectral capabilities of the up-coming X-ray calorimeters on board of
ASTRO-H \citep{Tak14,Kit14} and ATHENA \citep{Ett13}.  Spatially
unresolved rotation contributes, together with the turbulence, to the
emission line broadening: the present observational measures of the
X-ray emission line widths \citep{San13,Pin15}, even when combined
with measures of the flattening of the X-ray isophotes, leave ample
room for rotational motions \citep[][hereafter
  \citetalias{Bia13}]{Bia13}.  If the ICM rotates, its stability
properties are substantially modified \citep{Bal01,Nip10} and, in
particular, also the magnetorotational instability (MRI;
\citealt{BalH91}; see also \citealt{Vel59} and \citealt{Cha60}) can
play an important role \citep[][hereafter \citetalias{Nip14}]{Nip14}.

The linear evolution of local axisymmetric perturbations in a
stratified, rotating, radiatively cooling, weakly magnetized plasma is
determined by the dispersion relation derived in \citet[][hereafter
  \citetalias{Nip13}]{Nip13}. \citetalias{Nip14} studied the nature of
the instabilities of such a plasma in two illustrative cases,
generically representative of the physical conditions in galactic and
cluster atmospheres.  Here we focus on the cool cores of galaxy
clusters and we discuss the stability properties of rotating cool-core
models built on the basis of the work of \citetalias{Bia13}. The
present study extends the results of \citetalias{Nip14}, as here we
study the nature and linear growth rate of the plasma instabilities
throughout observationally motivated cool-core cluster models. In this
work we limit ourselves to axisymmetric perturbations, but the results
of \citetalias{Nip14} suggest that the study of non-axisymmetric
disturbances should lead to similar conclusions.  Though we consider
global cluster models, we study only the stability against local
perturbations, i.e. disturbances with sizes much smaller than the
characteristic scale-lengths of the system.  In addition to the local
instabilities found in the present study, the ICM could be subject to
global unstable modes \citep[e.g.][]{Lat12}, which by construction
would elude our analysis.

The paper is organized as follows.  The governing MHD equations are
given in Section~\ref{sec:mhd}, the unperturbed models are presented
in Section~\ref{sec:unp} and the dispersion relation obtained from the
linear-stability analysis is given in
Section~\ref{sec:lin}. Section~\ref{sec:res} describes the properties
of the local instabilities occurring in the cluster models and
Section~\ref{sec:con} concludes.

\section{Governing equations}
\label{sec:mhd}

A stratified, rotating, magnetized, dilute plasma in the presence of
thermal conduction and radiative cooling is governed by the following
MHD equations:
\begin{eqnarray}
&&{\de \rho \over \de t}+\grad\cdot(\rho\vv)=0,\label{eq:mass}\\
&&\rho\left[{\partial \vv \over \partial t}+\left(\vv\cdot\grad\right) \vv \right]=-\grad \left(p + {B^2 \over 8\pi}\right) -\rho\grad\Phi + {1 \over 4\pi}(\Bv \cdot \nabla)\Bv,\label{eq:mom}\\
&&{\de \Bv \over \de t} - \rot (\vv \times \Bv) = 0, \label{eq:indu}\\
&&{p\over \gamma-1}\left[{\partial \over \partial  t}+\vv\cdot\grad\right] \ln (p \rho^{-\gamma})=-\grad\cdot \Qv -\rho\L+\H,\label{eq:ene}
\end{eqnarray}
with the additional condition $\div \Bv = 0$.  Here $\rho$, $p$, $T$,
$\vv$ and $\Bv$ are, respectively, the density, pressure, temperature,
velocity and magnetic field of the fluid, $\Phi$ is the external
gravitational potential (we neglect self-gravity), $\gamma=5/3$ is the
adiabatic index, $\H$ is the heating rate per unit volume,
$\L=\L(T,\rho)$ is the radiative energy loss per unit mass of fluid,
and
\begin{eqnarray}
&&
\Qv = -\frac{\chi \Bv\left(\Bv\cdot\nabla\right)T}{ B^2}
\label{eq:heatflux}
\end{eqnarray}
is the conductive heat flux, where $\chi \equiv \kappa T^{5/2}$ is
\citet{Spi62} electron conductivity with
$\kappa\simeq{1.84\times10^{-5}(\ln{\Lambda})}^{-1}\allowbreak \erg
\sminusone \cmminusone \kelvin^{-7/2}$ and $\ln\Lambda$ is the Coulomb
logarithm.  Neglecting the weak temperature and density dependence of
$\ln\Lambda$, here we fix $\ln \Lambda=30$, so $\kappa$ is a constant
and $\chi=\chi(T)\propto T^{5/2}$.  As we consider rotating fluids, we
work in cylindrical coordinates ($R$, $\phi$, $z$), where $R=0$ is the
rotation axis.  The governing equations~(\ref{eq:mass}-\ref{eq:ene})
are given explicitly in cylindrical coordinates in
Appendix~\ref{app:cyl}.

The expression~(\ref{eq:heatflux}) for $\Qv$ accounts for the fact
that in a dilute magnetized plasma heat is significantly transported
by electrons only along the magnetic field lines. For simplicity, even
if the medium is magnetized, we have assumed that the pressure is
isotropic, neglecting the fact that momentum transport is anisotropic
in the presence of a magnetic field (i.e. the so-called Braginskii
viscosity; \citealt{Bra65}).  Though this approximation is not
rigorously justified, we adopt it in the working hypothesis that
anisotropic pressure is not the crucial factor in determining the
dominant instabilities of a rotating ICM. The importance of the
effects of the Braginskii viscosity on plasma stability is debated
\citep{Kun11,Lat12,Par12}, so the limits of this approximation must be
taken into account when interpreting the results of the present work.

The term $\H$ in equation~(\ref{eq:ene}) represents an unspecified
source of heating that is expected to counteract radiative cooling and
prevent a cooling catastrophe. There is general consensus that heat
conduction cannot be entirely responsible for halting cooling flows
\citep[e.g.][]{Par09} and that other sources of heating, such as
feedback from the central active galactic nucleus (AGN), must be at
work to balance cooling in a time-averaged sense. In the absence of
physically motivated and analytically tractable models of such a
heating term, here we just consider a toy model in which, for a given
cluster model, $\H$ is a fixed position-dependent function \citep[for
  a discussion see][]{Mcc12}: the main results of this work are
independent of this approximation, though very crude.

\section{Cool-core cluster models}
\label{sec:unp}

\begin{figure}
  \centerline{\includegraphics{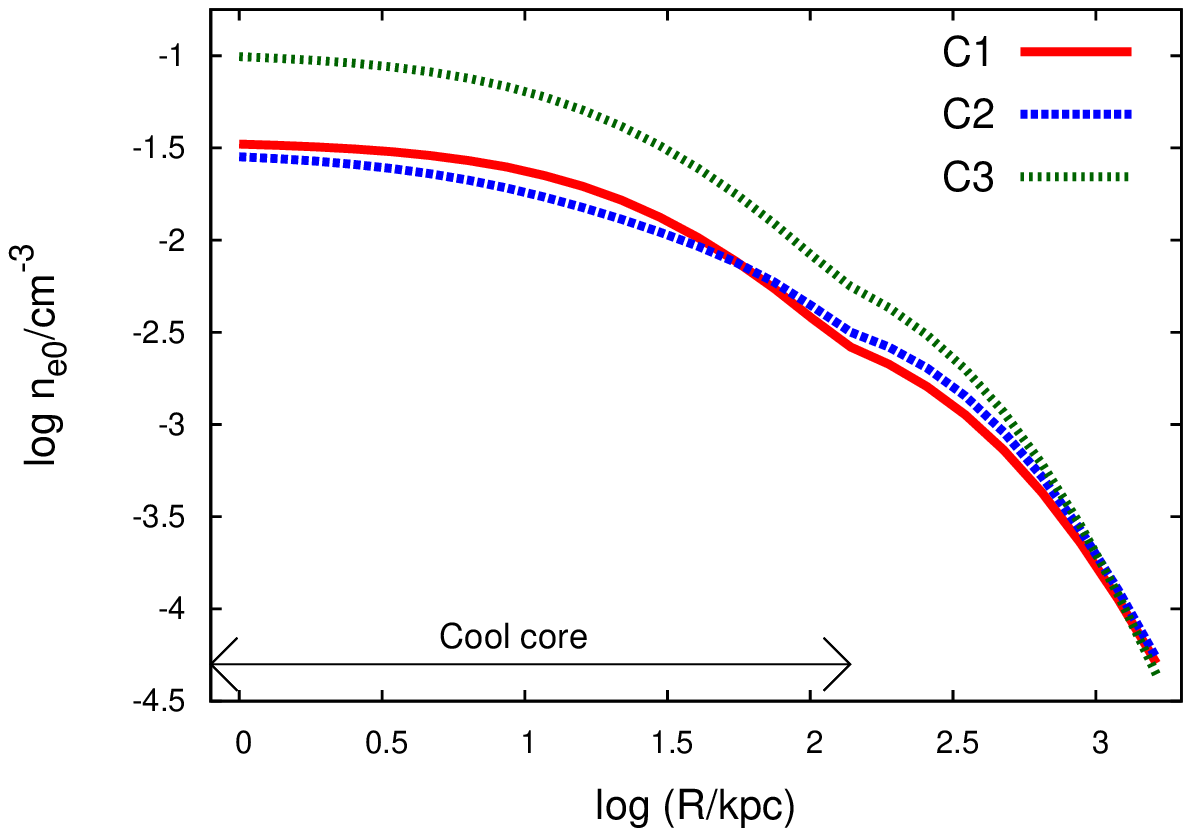}}
  \caption{ICM electron number density profiles in the equatorial plane
    of the rotating cool-core cluster models C1, C2 and C3 (see
    Section~\ref{sec:unp}).}
\label{fig:nezero}
  \centerline{\includegraphics{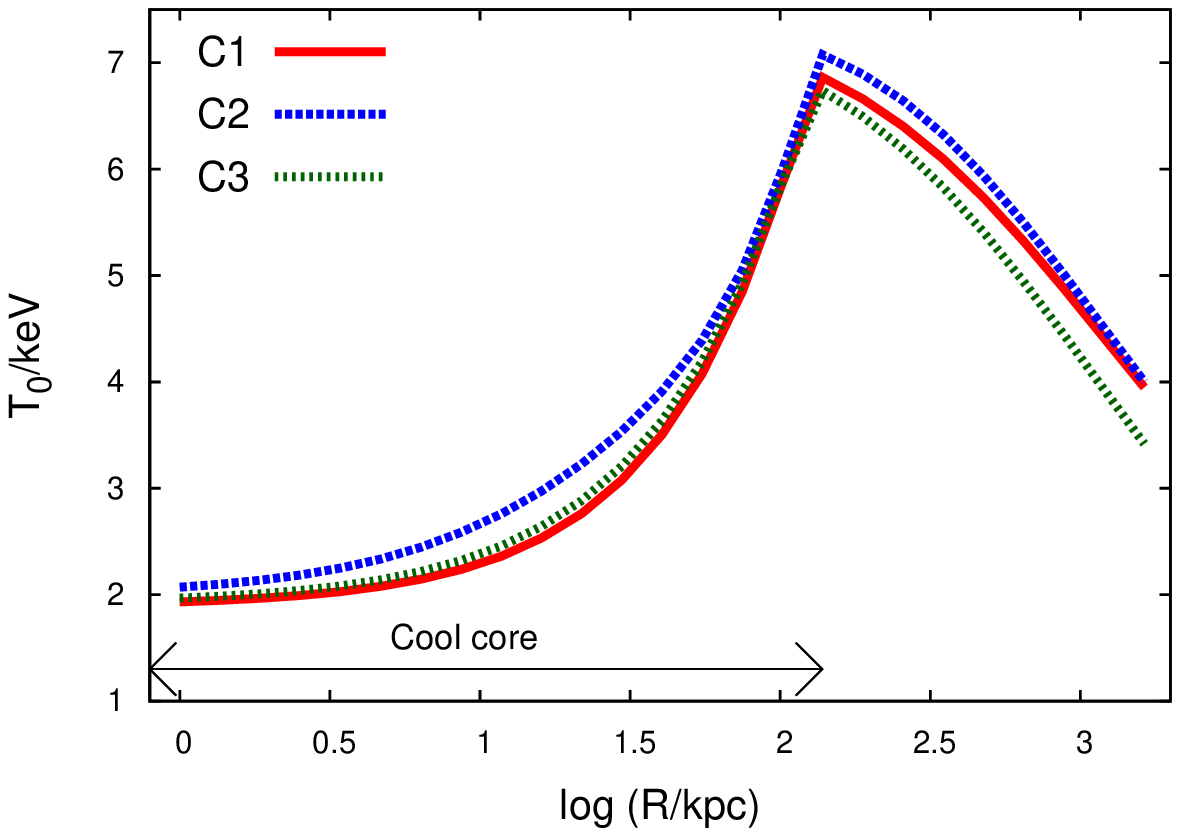}}
  \caption{ICM temperature profiles in the equatorial plane of the same
    models as in Fig.~\ref{fig:nezero}.}
\label{fig:tzero}
\end{figure}

\begin{figure}
  \centerline{\includegraphics{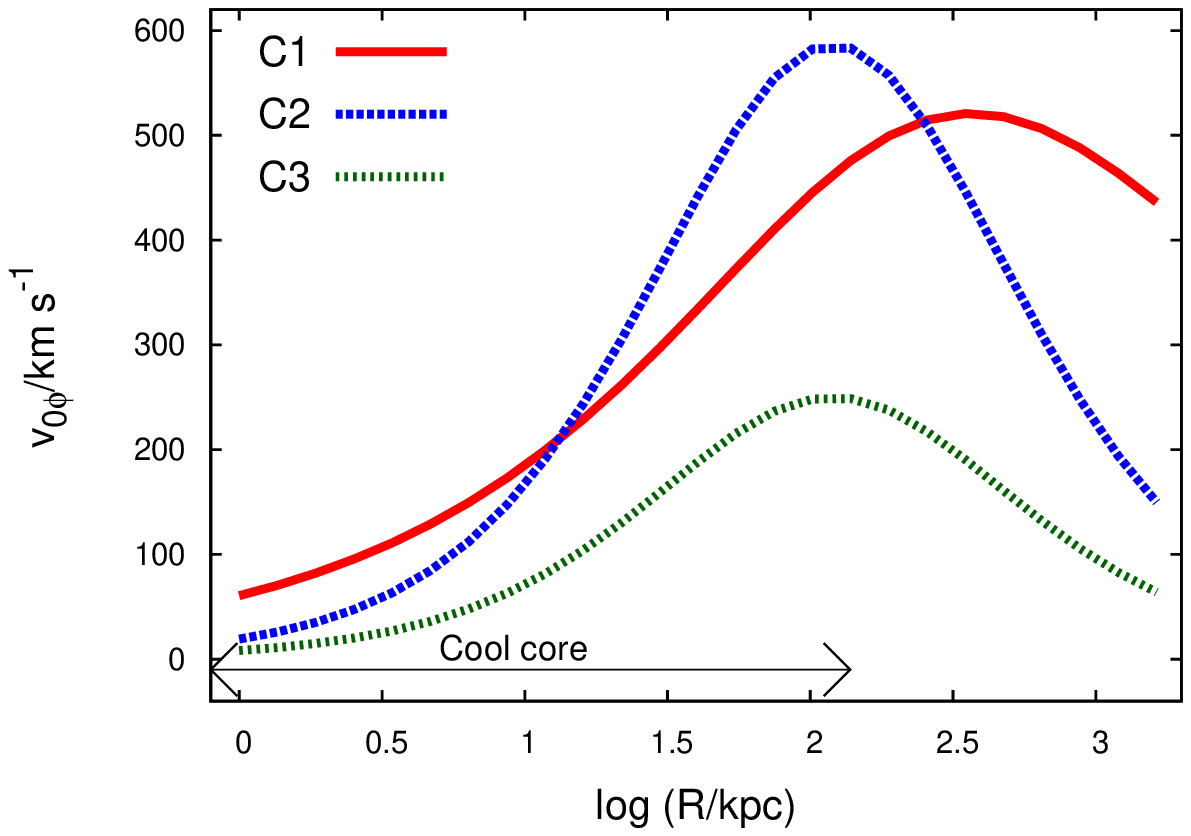}}
  \caption{ICM rotation speed as a function of radius for the same
    models as in Fig.~\ref{fig:nezero}.}
\label{fig:vphi}
  \centerline{\includegraphics{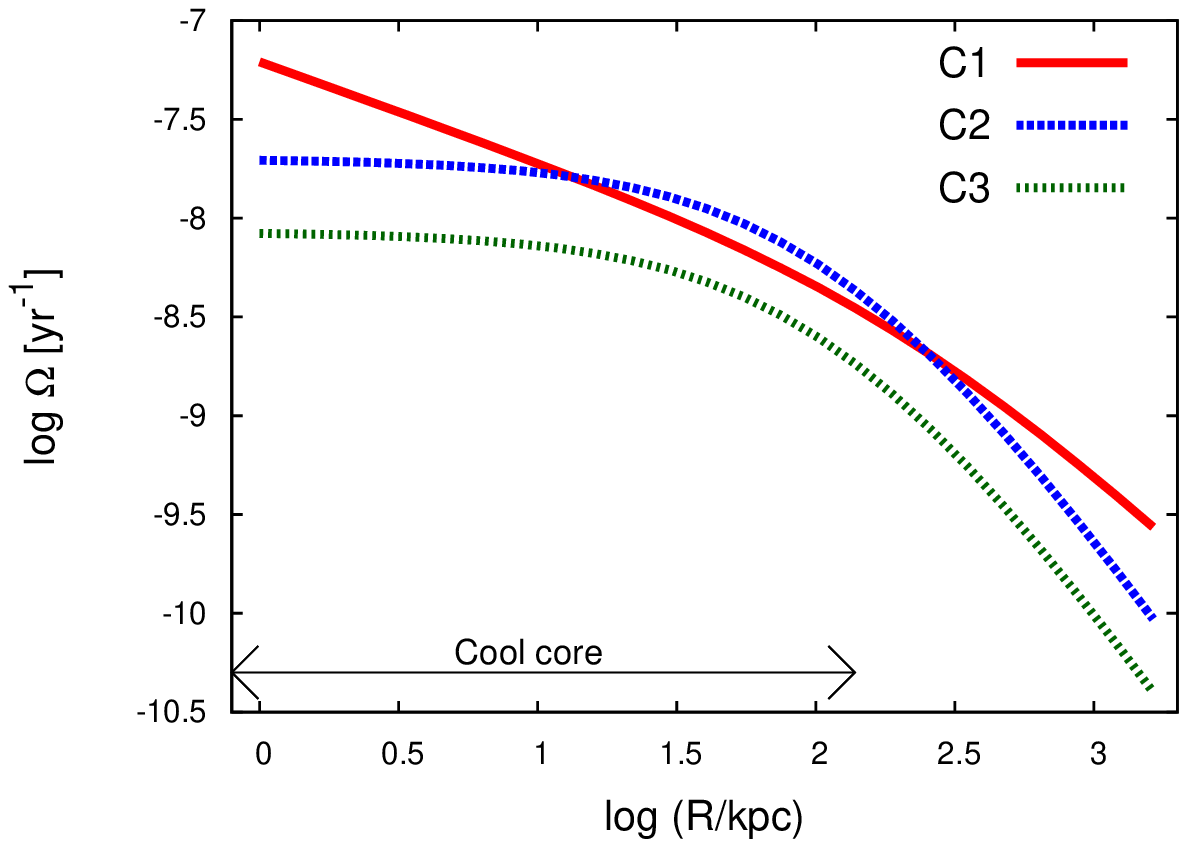}}
  \caption{ICM angular velocity as a function of radius for the same
    models as in Fig.~\ref{fig:nezero}.}
\label{fig:omega}
\end{figure}

\begin{figure}
  \centerline{\includegraphics{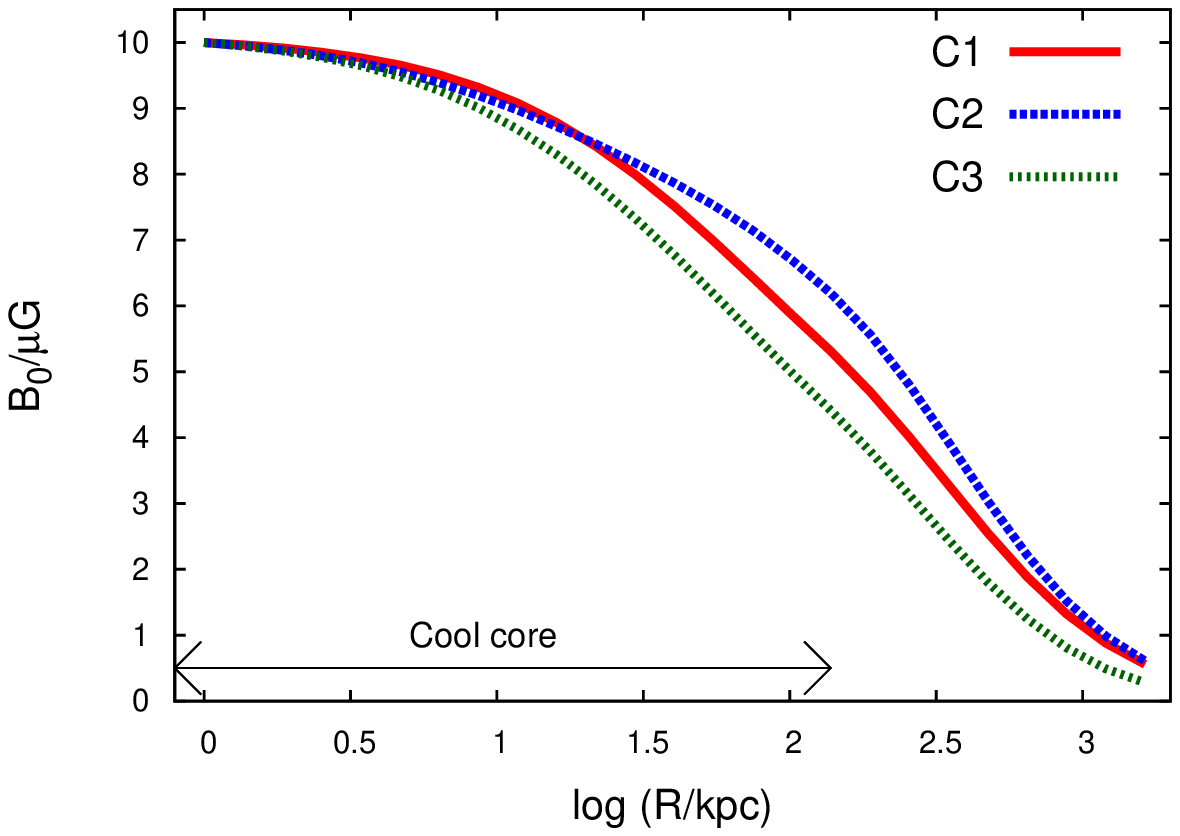}}
  \caption{ICM magnetic field modulus as a function of radius in the
    equatorial plane of the same models as in Fig.~\ref{fig:nezero}.}
\label{fig:bzero}
\end{figure}

We consider here three rotating cool-core cluster models (named C1, C2
and C3), which are slightly modified versions of those presented in
\citetalias{Bia13} (indicated as CC1, CC2 and CC3 in that paper).  In
all these models, which represent massive galaxy clusters, the total
gravitational potential $\Phi$ is given by a spherical\footnote{ In a
  more realistic model the total gravitational potential would be
  axisymmetric or triaxial. Though idealized, a model with spherical
  potential is an interesting limiting case because the flattening of
  the ICM distribution is entirely due to rotation.}
Navarro-Frenk-White (NFW) model \citep{Nav95} with virial mass
$M_{200}=10^{15}\Msun$, scale radius $\rs=519\kpc$ and virial radius
$r_{200}=2066\kpc$. The plasma rotates differentially with azimuthal
velocity $\vzerophi=\Omega R$, where $\Omega=\Omega(R)$ is the angular
velocity and the rotation law is different in the three models (see
section 2.3 of \citetalias{Bia13}). The other velocity components are
null.  Assuming that the ICM magnetic field is dynamically
unimportant, we construct the unperturbed pressure
$\pzero=\pzero(R,z)$, density $\rhozero=\rhozero(R,z)$ and temperature
$\Tzero=\Tzero(R,z)$ fields as axisymmetric stationary solutions of
equations~(\ref{eq:mass}-\ref{eq:mom}) with $\Bv=0$. As $\Omega$ does
not depend on $z$, the gas distribution is barotropic ($\rhozero$,
$\pzero$ and $\Tzero$ are stratified on the surfaces of constant
effective potential). In particular, the plasma is modeled as an
equilibrium two-component composite polytrope with outer polytropic
index $\gammaouttil=1.14$ and inner polytropic index $\gammaintil<1$,
so the ICM temperature decreases outwards in the outer regions and
increases outwards in the inner regions (cool core). The parameters of
models C1, C2 and C3 are, respectively, the same as those of models
CC1, CC2 and CC3 of \citetalias{Bia13}, with the only exception of the
inner polytropic index $\gammaintil$, which is $\gammaintil=0.49$ in
model C1, $\gammaintil=0.43$ in model C2, $\gammaintil=0.56$ in model
C3 (these values, smaller than in \citetalias{Bia13}, lead to higher
central densities and lower central temperatures).  The plasma
temperature is in the range $2-7 \keV$ and the size of the cool core
is about $140\kpc$ in the equatorial plane. As in \citetalias{Bia13},
we assume that the ICM metallicity is everywhere $3/10$ of the solar
metallicity.  Figs.~\ref{fig:nezero}-\ref{fig:omega} plot the radial
profiles, in the equatorial plane, of electron number density
$\nezero$, temperature $\Tzero$, rotation speed $\vzerophi$ and
angular velocity $\Omega$.  As discussed in \citetalias{Bia13}, the
above models of rotating clusters are realistic, in the sense that
they are consistent with the current measurements of the ellipticity
of the X-ray isophotes and of the width of the X-ray emission
lines. Additional constraints on the rotation of the ICM derive from
the so-called {\it hydrostatic mass bias}, that is the discrepancy
between cluster mass estimates from gravitational lensing and those
based on the assumption that the ICM is in hydrostatic equilibrium. In
Appendix~\ref{app:bias} we show that models C1, C2 and C3 are
realistic also in this respect.

Given a cluster model, we assume that the heating term $\H=\H(R,z)$
appearing in the energy equation~(\ref{eq:ene}) is such that
$\H-\rho_0\L(\Tzero,\rhozero)-\nabla\cdot\Qvzero=0$, where $\Qvzero$
is the unperturbed conductive heat flux. We remark that in our models
$\H$ is not a function of the hydrodynamic variables, but depends
explicitly on the coordinates $R$ and $z$. To complete our cluster
models we must specify the properties of the unperturbed magnetic
field $\Bvzero$, which, though not relevant to the equilibrium
configuration (because $\beta\equiv 8\pi \pzero/\Bzero^2\gg 1$),
influences the stability of the ICM. We do not attempt a full
characterization of the magnetic field throughout the global cluster
model. For simplicity, in this work we focus only on the equatorial
plane $z=0$, and we specify locally $\Bvzero$ as a function of
cylindrical radius $R$.  The stationary unperturbed magnetic field
must satisfy \citet{Fer37} isorotation law
$\Bvzero\cdot\nabla\Omega=0$, so $\BzeroR=0$ in the considered
barotropic distributions.  For each cluster model, we consider two
cases: one in which the only non-vanishing component of the
unperturbed magnetic field is $\Bzeroz$ (so $\Bzerophi=0$) and another
in which both $\Bzeroz$ and $\Bzerophi$ are non-vanishing
($\Bzerophi/\Bzeroz=10$). In all cases, given the unperturbed pressure
distribution $\pzero$, the unperturbed magnetic field modulus is fixed
by imposing a constant value of $\beta$ such that at the centre
$\Bzero=10\mu\G$ (consistent with observational estimates;
e.g. \citealt{Bru13}), i.e.  $\beta=29.3$ for model C1, $\beta=26.8$
for model C2 and $\beta=89$ for model C3. The profiles of the magnetic
field modulus for the three models are shown in Fig.~\ref{fig:bzero}.
We note that in the equatorial plane the unperturbed conductive heat
flux $\Qvzero$ vanishes, because at $z=0$ the magnetic field lines are
isothermal (the only non vanishing component of the temperature
gradient is the radial one, which, combined with $\BzeroR=0$, implies
$\Qvzero=0$), so the condition of energy balance is simply
$\H=\rho_0\L(\Tzero,\rhozero)$.

\section{Linear stability analysis}
\label{sec:lin}

The stability analysis is performed by linearizing the system
(\ref{eq:mass}-\ref{eq:ene}) with local axisymmetric Eulerian
perturbations of the form $f\e^{-\i\omega t + \i\kR R + \i\kz z}$ with
$|f|\ll|\fzero|$, where $\fzero$ is the unperturbed quantity, $\omega$
is the perturbation frequency, and $\kR$ and $\kz$ are, respectively,
the radial and vertical components of the perturbation wave-vector
$\kv$.  In the short wave-length and low frequency approximation, the
resulting dispersion relation for $n\equiv-\i\omega$ is
(\citetalias{Nip13})
\begin{eqnarray}
\label{eq:disprel}
&& n^5 + \omegad n^4 + \left[\omegaBVsq + \omegarotsq + 2 \omegaAsq\right]n^3 + \left[(\omegarotsq + 2 \omegaAsq)\omegad + \omegaAsq\omegacmag \right]n^2 \nonumber\\
&&\qquad + \omegaAsq \left( \omegaAsq 
+ \omegaBVsq + \omegarotsq - 4\Omega^2 \frac{\kz^2}{k^2}+\omegacphisq\right)n \nonumber \\ 
&&\qquad+ \omegaAsq\left[ \left(\omegaAsq + \omegarotsq - 4\Omega^2\frac{\kz^2}{k^2} \right)\omegad  + \omegaAsq\omegacmag \right] = 0,
\end{eqnarray}
where the quantities indicated with subscripted $\omega$ are
characteristic frequencies related to local properties of the
unperturbed distribution and of the wave-vector. Specifically,
$\omegad\equiv\omegath+\omegaca$, where $\omegath$ is the TI frequency
and $\omegaca$ is the anisotropic thermal conduction frequency,
$\omegaBV$ is the Brunt-V\"ais\"al\"a frequency, $\omegarot$ is the
angular momentum gradient frequency, $\omegaA$ is the Alfv\'en
frequency, and $\omegacmag$ and $\omegacphi$ are two other frequencies
associated with thermal conduction mediated by the magnetic field (all
the definitions are given in \citetalias{Nip13}). The radiative energy
loss per unit mass of fluid $\L(\rhozero,\Tzero)$, appearing in
$\omegath$ (see \citetalias{Nip13}), is computed using the collisional
ionization equilibrium cooling function of \citet{Sut93}.  In the
dispersion relation~(\ref{eq:disprel}) there are no terms related to
the heating source $\H$ appearing in equation~(\ref{eq:ene}), because
we are considering Eulerian perturbations at a fixed point in space
and $\H$ is a function of position, but not of the hydrodynamic
variables. For fixed wave-vector $\kv$, given a cool-core cluster
model (see Section~\ref{sec:unp}), the coefficients of
equation~(\ref{eq:disprel}) are fully determined, so the problem of
the local linear stability of the plasma is reduced to finding the
roots of a polynomial equation.

\section{Results}
\label{sec:res}

For each of the considered cluster models, at different radii $R$ in
the equatorial plane, for given wave-vector $\kv$ we numerically
computed the solutions of the dispersion relation~(\ref{eq:disprel})
using the {\idl} (Interactive Data Language) routine {\fzroots} (as in
\citetalias{Nip14}). Depending on the signs of their real ($\Rn$) and
imaginary ($\In$) parts, the solutions are classified as stable
($\Rn\leq 0$), overstable ($\Rn> 0$, $\In\neq0$) or monotonically
unstable ($\Rn> 0$, $\In=0$) modes. As overstable linear disturbances
are unlikely to enter the non-linear regime \citep{Mal87,Bin09}, here
we focus only on the more interesting monotonically unstable linear
modes. For each model we compute, as a function of radius $R$, the
growth rate $\ninst$ of the fastest-growing monotonically unstable
mode. At each radius, $\ninst$ is calculated considering modes with
wave-number $k=\sqrt{\kR^2+\kz^2}$ in the range $10< kh<1000$, where
$h\equiv (|\nabla p|/p)^{-1}$ is the pressure scale-height, which
depends on $R$: this guarantees that the perturbation wave-length is
short with respect to the characteristic scales of the system at all
radii.  The radial-to-vertical wave-number ratio $\kR/\kz $ is allowed
to vary in the range $-25\leq \kR/\kz\leq 25$.

\begin{figure}
  \centerline{\includegraphics{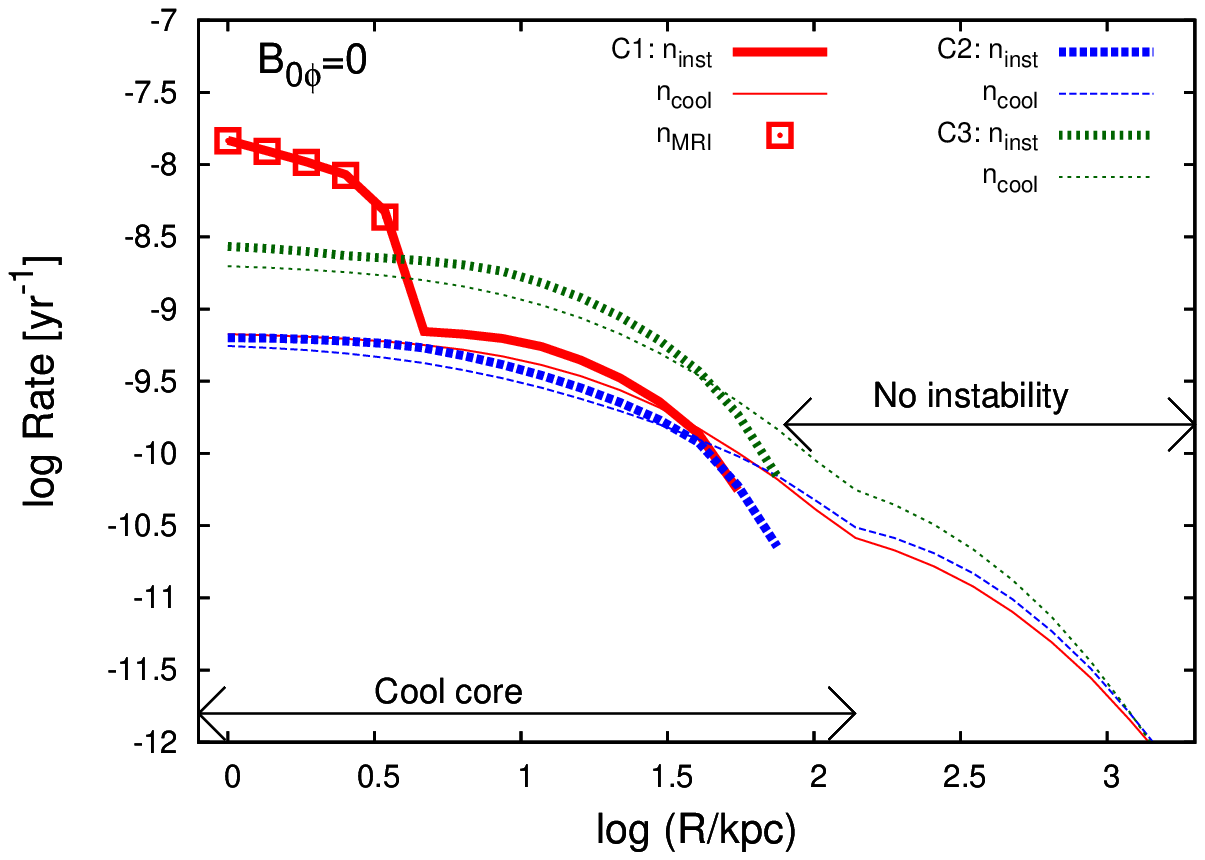}}
  \caption{Maximum instability growth rate $\ninst$, cooling rate
    $\ncool$, and MRI growth rate $\nMRI$ as functions of radius in
    the equatorial plane of the same models as in
    Fig.~\ref{fig:nezero}. Here $\Bzerophi=\BzeroR=0$ and
    $\Bzeroz\neq0$. At $R \gtrsim 80\kpc$ there are no monotonically
    unstable modes, at least in the explored wave-vector space.}
\label{fig:rate}
  \centerline{\includegraphics{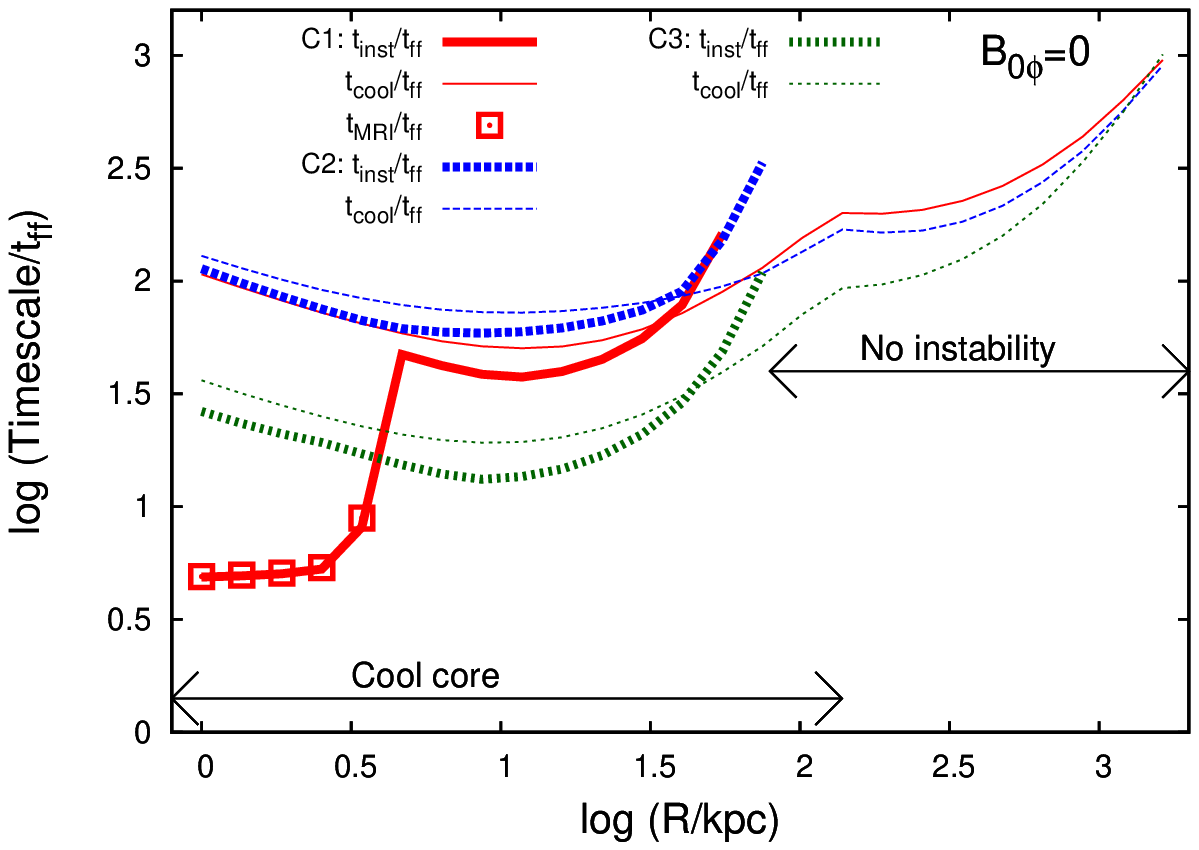}}
  \caption{Instability ($\tinst$), cooling ($\tcool$) and MRI
    ($\tMRI$) timescales normalized to the free-fall time $\tff$ as
    functions of radius in the equatorial plane of the same models as
    in Fig.~\ref{fig:rate}.}
\label{fig:time}
\end{figure}

As discussed in \citetalias{Nip13} and \citetalias{Nip14}, in general
the instabilities of rotating, weakly magnetized, radiatively cooling
plasmas arise from combinations of TI, MTI, HBI and MRI.  In order to
understand the physical consequences of the instability it is
important to determine the nature of the dominant unstable modes.
\citetalias{Nip14} have shown that, though the different modes are
entangled, it is possible to identify the different branches of the
solutions in terms of combinations of well-known modes (TI, MRI, MTI,
HBI, rotation, buoyancy and Alfv\'en modes). In practice, this is done
by inspecting the behaviour and the growth rates of the individual
branches of the solutions as functions of the wave-number and
comparing the results with those obtained for simpler configurations
(for instance in the absence of radiative cooling, rotation or
magnetic field). Applying this technique to the cluster models here
considered, we have been able to identify the nature of the dominant
instabilities at different radii.

It is useful to compare the maximum instability growth rate $\ninst$
with the characteristic rates of radiative cooling and of the MRI.
Therefore, we define the cooling rate
\begin{equation}
\ncool\equiv\frac{\gamma-1}{\gamma}\frac{\mu\mp\L(\Tzero,\rhozero)}{\kb\Tzero},
\end{equation}
which for each unperturbed model is a function of position (here
$\mu=0.59$ is the mean mass per particle in units of the proton mass
$\mp$).  The MRI growth rate $\nMRI$ is the maximum growth rate of
monotonically unstable modes that are solutions of the dispersion
relation \citep{BalH91}
\begin{eqnarray}
\label{eq:disprelmri}
&& n^4 
+  \left[\omegaBVsq + \omegarotsq + 2 \omegaAsq\right]n^2
+ \omegaAsq \left( \omegaAsq + \omegaBVsq + \omegarotsq - 4\Omega^2 \frac{\kz^2}{k^2}\right) = 0,
\end{eqnarray}
which is obtained from the more general dispersion
relation~(\ref{eq:disprel}) in the limit of absence of radiative
cooling and thermal conduction ($\omegad=0$, $\omegacmag=0$,
$\omegacphi=0$). As well known, the TI growth rate $|\omegath|$ is of
the order of, but typically somewhat higher than the cooling rate
$\ncool$, which is at the basis of the existence of the TI in
astrophysical plasmas \citep{Fie65}. The MRI growth rate $\nMRI$ is
proportional to the plasma angular velocity $\Omega$ \citep{BalH91}.

As mentioned in Section~\ref{sec:unp}, for each cluster model, we
consider two cases: one in which $\Bzerophi=0$ and another in which
$\Bzerophi/\Bzeroz=10$ at all radii. The only difference between these
two families of model is the value of the ratio $\Bzerophi/\Bzeroz$:
all the other properties are the same, including the radial profile of
the magnetic field modulus, which is shown in Fig.~\ref{fig:bzero}.

\subsection{Models with vanishing azimuthal magnetic field ($\Bzerophi=0$)}

Fig.~\ref{fig:rate} plots, as a function of radius, the maximum growth
rate $\ninst$ in the equatorial plane of models C1, C2 and C3 when
$\Bzerophi=0$. For comparison, the cooling ($\ncool$) and MRI
($\nMRI$) rates are also plotted for the same models. All models are
unstable in their cool core, while no monotonically unstable modes are
found at $R\gtrsim 80\kpc$. Model C1 is unstable for MRI in the inner
$\sim 10\kpc$ (instability timescale $\sim10^8\yr$), while at larger
radii the system is unstable for TI (instability timescale $\sim
10^9\yr$, of the order of the cooling time).  Models C2 and C3 are not
unstable for MRI, but present TI modes throughout the cool core with
instability timescales between $5\times10^8$ and $5\times 10^9\yr$.
Together with the absolute timescales, it is also useful to consider
the timescales normalized to the local dynamical (``free-fall'') time
$\tff\equiv[2 r / g(r)]^{1/2}$, where $g=\d \Phi/\d r$ is the modulus
of the gravitational force and $r$ is the radial spherical coordinate.
Fig.~\ref{fig:time} plots, as functions of radius, the instability
($\tinst\equiv\ninst^{-1}$), cooling ($\tcool\equiv\ncool^{-1}$) and
MRI ($\tMRI\equiv \nMRI^{-1}$) timescales normalized to $\tff$. It is
apparent that, when present, the MRI grows on timescales as short as
$5\tff$, while the TI timescale is typically $10-100\tff$.

Though all models satisfy at all radii the necessary condition to have
MRI ($\Bzeroz\neq0$ and $\d \Omega /\d R <0$; see
Fig.~\ref{fig:omega}), MRI modes are present only in the inner core of
model C1.  We recall that the MRI occurs only for wave-numbers smaller
than a critical wave-number $\kMRI$, which, with the exception of the
centre of model C1, is outside the explored wave-number interval (in
other words, short wave-length perturbations are not unstable for MRI
in these cases). Neither the HBI nor the MTI contributes to the
unstable modes: these instabilities are related to anisotropic thermal
conduction and can occur when the temperature either increases (HBI)
or decreases (MTI) outwards.  In the cool core the temperature
increases outwards, but the magnetic field lines are isothermal so no
HBI is expected.  Out of the cool core the temperature decreases
outwards, so in principle the MTI might occur, but in fact no MTI
modes are found in the explored wave-vector space. Similarly to the
MRI, the MTI occurs only for wave-numbers smaller than a critical
wave-number $\kMTI$ \citep{Bal00}, which is out of the explored
wave-number interval (as it happens for the MRI, short wave-length
disturbances are not unstable for MTI in these cases).

\subsection{Models with dominant azimuthal magnetic field ($\Bzerophi/\Bzeroz=10$)}

\begin{figure}
  \centerline{\includegraphics{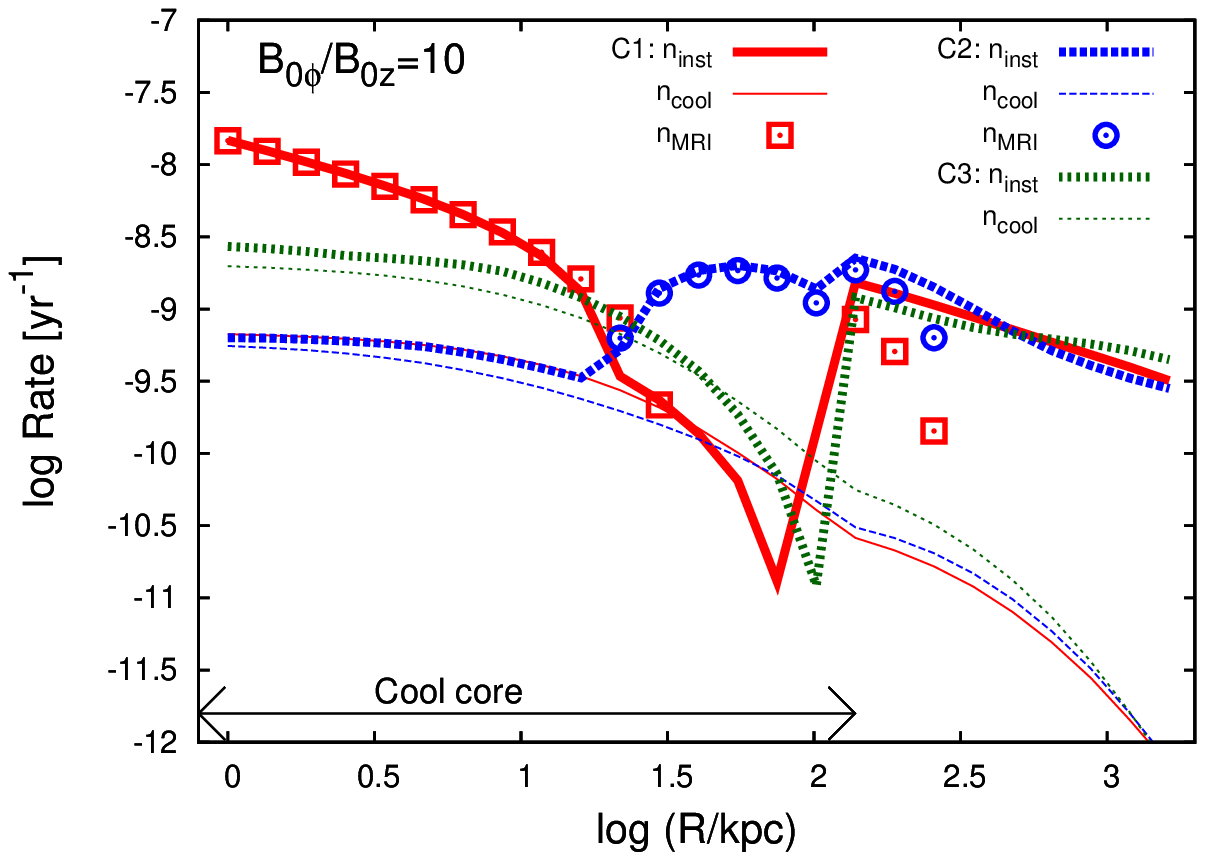}}
  \caption{ Maximum instability growth rate $\ninst$, cooling rate
    $\ncool$, and MRI growth rate $\nMRI$ as functions of radius in
    the equatorial plane of the same models as in
    Fig.~\ref{fig:nezero}. Here $\BzeroR=0$, $\Bzeroz\neq0$ and
    $\Bzerophi/\Bzeroz=10$.}
\label{fig:rate_az}
  \centerline{\includegraphics{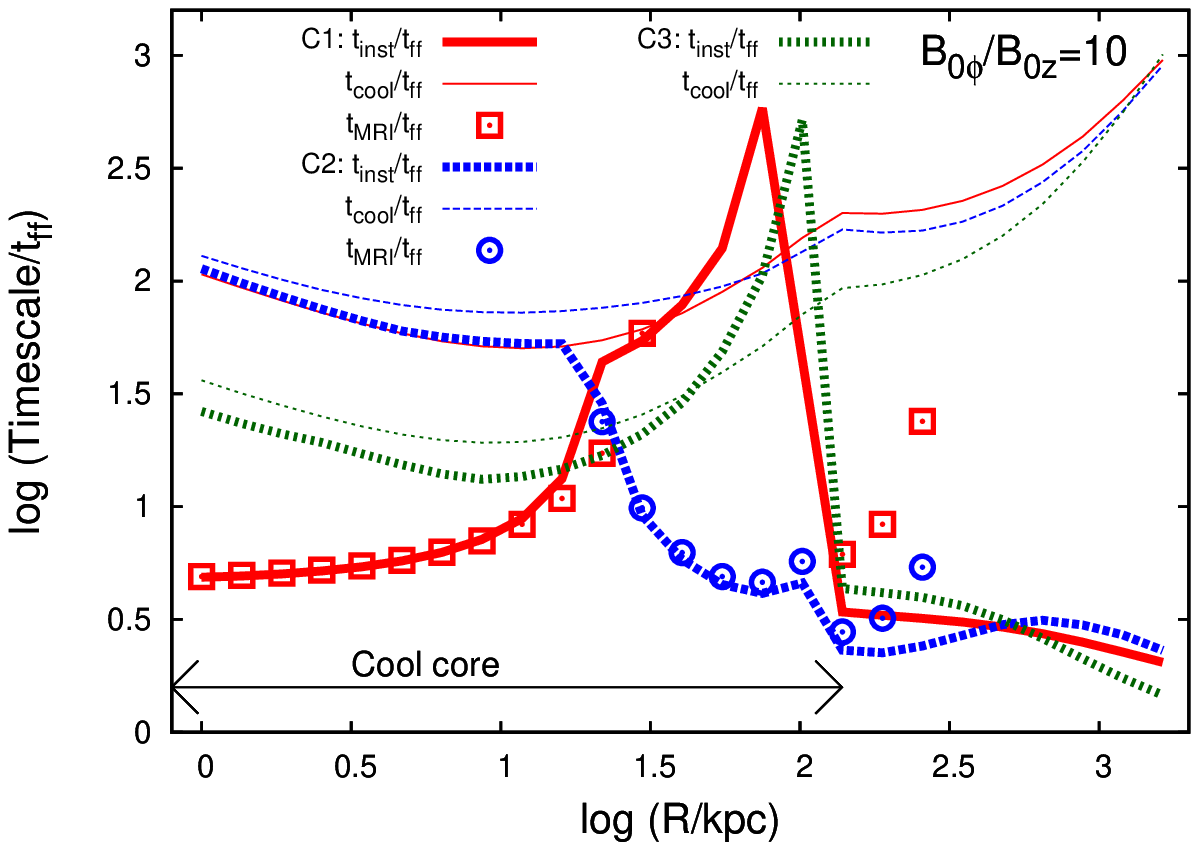}}
  \caption{Instability ($\tinst$), cooling ($\tcool$) and MRI
    ($\tMRI$) timescales normalized to the free-fall time $\tff$ as
    functions of radius in the equatorial plane of the same models as
    in Fig.~\ref{fig:rate_az}.}
\label{fig:time_az}
\end{figure}

Figs.~\ref{fig:rate_az}-\ref{fig:time_az} plot for models C1, C2 and
C3 the same quantities as Figs.~\ref{fig:rate}-\ref{fig:time}, but
under the assumption that the azimuthal magnetic field component is
dominant ($\Bzerophi/\Bzeroz=10$).  In this case all models present
monotonically unstable modes out to large radii ($\sim\Mpc$). At all
radii the unstable modes of model C1 have growth rate much higher than
the cooling rate: the MRI dominates throughout the cool core, while in
the cluster outskirts the unstable modes are driven by the MTI. Model
C2 presents TI modes in the inner parts of the cool core, MRI modes in
the intermediate radial range ($10-100\kpc$) and MTI modes in the
outskirts. Model C3, which has the lowest rotation speed at all radii
(Fig.~\ref{fig:vphi}) is stable against MRI even when $\Bzerophi$ is
dominant: its unstable modes are driven by TI in the cool core and by
MTI in the outer regions where the temperature gradient is negative.
In the explored models, both the MRI and the MTI modes have remarkably
short timescales, of the order of a few local dynamical times. In
absolute terms the MRI modes are fastest with growth times as short as
$\sim10^8\yr$ in the core.

Comparing Figs.~\ref{fig:rate_az}-\ref{fig:time_az} with
Figs.~\ref{fig:rate}-\ref{fig:time} it is apparent that the
instabilities are stronger and more widespread when the magnetic field
is dominated by the azimuthal component. The driving factor in
determining these differences is not the presence of an azimuthal
magnetic field component, but the fact that when $\Bzerophi$ is
dominant, the vertical component of the magnetic field is weaker (at
fixed $\Bzero$). As pointed out above, the MRI and the MTI occur only
for wave-numbers smaller than critical values ($\kMRI$ and $\kMTI$,
respectively). Both $\kMRI$ and $\kMTI$ are inversely proportional to
the magnetic field component coupled with the wave-vector
\citep{BalH91,Bal00}, which in the present case is just
$\Bzeroz$. When the azimuthal magnetic field is dominant, the vertical
magnetic field is sufficiently weak to trigger the MRI throughout the
cores of models C1 and C2, and the MTI in the outer regions of all
models.  It is also the case that $\kMRI$ increases for increasing
angular velocity gradient, which explains why the MRI manifests itself
in models C1 and C2, which have stronger angular velocity gradients
than model C3 (see Fig.~\ref{fig:omega}).  When present, both the MRI
and the MTI have growth rates much higher than the local cooling rate:
the fastest instability is the MRI, which acts on timescales
$10^8-10^9\yr$.

\section{Conclusions}
\label{sec:con}

Building upon the results of \citetalias{Nip13}, \citetalias{Nip14}
and \citetalias{Bia13}, in this paper we have studied local
instabilities arising from axisymmetric perturbations in rotating
cool-core clusters.  For the sake of simplicity we limited ourselves
to somewhat idealized cluster models. Notwithstanding these
limitations, our calculations should catch the essential properties of
more realistic cluster models that could be built in the near future,
especially when more observational constraints on the ICM rotation are
available, thanks to the next-generation X-ray missions such as
ASTRO-H  and ATHENA.

The results of the linear stability analysis of the rotating cool-core
cluster models here presented show that, depending on the ICM
properties, TI, MTI and MRI can occur at different radii in the
equatorial plane (where, by construction, our models exclude the HBI,
which however can play a role in more general configurations).  What
are the astrophysical consequences of these instabilities?

In principle, monotonically unstable TI modes can lead to local
condensation of cold gas out of the hot plasma and therefore to a
multiphase ICM. However, given that the growth rate of TIs is of the
order of the cooling rate, the evolution of thermally unstable modes
depend on the details of the balance of heating and cooling in the
unperturbed fluid.  While heat conduction always tend to damp thermal
perturbations, other heating mechanisms such as AGN feedback have more
complex effects and can in some cases even favour cooling
instabilities \citep[e.g.][]{Cio07,Gas12}. Therefore, it is not
possible to draw robust conclusions on condensational modes based on
the models here considered, in which we have adopted a very simple
{\it ad hoc} heating source to balance radiative losses in the
background plasma. Multiphase plasma is detected in cluster cool cores
\citep{Hec81}, but the origin of the colder medium is still
debated. The observed correlation between presence of multiphase gas
and short cooling time (low entropy) in cluster cores
\citep[e.g.][]{Voi15} does not necessarily imply that the cold gas is
produced by TI \citep{Nip04}.

As far as the MTI is concerned, our results confirm previous findings
that the MTI can be the dominant instability out of the cluster core.
As in non-rotating ICM models, the MTI can affect the outer regions of
the cluster (where the temperature decreases outwards) on times scales
as short as a few local dynamical times, potentially driving strong
turbulence in the non-linear regime \citep{Mcc11}. We recall that our
models do not account self-consistently for anisotropic momentum
transport, which in principle can affect the behaviour of the
MTI. However, it has been shown that, at least in non-rotating
systems, the effects of the Braginskii viscosity, which can be
significant for the HBI \citep{Lat12}, are almost negligible for the
MTI \citep{Kun12}.

The most interesting instabilities among those found in this work are
the MRI modes in the cool
cores. Figs.~\ref{fig:rate}-\ref{fig:time_az} show that the MRI rates
are remarkably high, with associated timescales $10^8-10^9\yr$ (a few
local dynamical times): this suggest that the MRI can have an
important impact on cool cores. The MRI timescales are substantially
shorter than the cooling time, so this conclusion is robust against
uncertainties on the balance of heating and cooling in the core.  The
Braginskii viscosity, which is neglected in our calculations, could
even enhance the growth rates of these dominant MRI modes
\citep{Qua02,Bal04}.  Studies of weakly magnetized accretion discs
\citep{Bal94} have shown that the MRI is expected to drive turbulence,
with associated energy and angular momentum transport. These
MRI-driven dissipative processes can contribute substantially to the
dynamics and evolution of cluster cool cores, if they rotate
significantly. The MRI could be the engine of turbulent heating, which
is believed to be an efficient mechanism to halt cooling flows in
galaxy clusters \citep{Zhu14}. A natural development of the present
work will be a quantitative estimate of the effect of the MRI on the
energy balance and accretion rate of the plasma in cluster cool
cores.\\

We acknowledge helpful discussions with Steven Balbus and Richard
Lovelace.  We thank the two anonymous referees for useful comments. CN
has been supported by PRIN MIUR 2010-2011, project ``The Chemical and
Dynamical Evolution of the Milky Way and Local Group Galaxies'',
prot. 2010LY5N2T.

\appendix

\section{Governing equations in cylindrical coordinates}
\label{app:cyl}

In cylindrical coordinates the governing
equations~(\ref{eq:mass}-\ref{eq:ene}) read
\begin{eqnarray}
&& \frac{\de \rho}{\de t} + \frac{1}{R}\frac{\de R\rho \vR}{\de R} + \frac{1}{R}\frac{\de \rho \vphi}{\de \phi} + \frac{\de \rho \vz}{\de z} = 0 \label{eq:model_cyl1}, \\
&& \frac{\de \vR}{\de t} + \vv\cdot\nabla\vR - \frac{\vphi^2}{R} = -\frac{1}{\rho}\frac{\de p}{\de R} - \frac{\de \Phi}{\de R} 
+ \frac{1}{4\pi\rho}\left( \Bv\cdot\nabla\BR - \frac{\Bphi^2}{R}\right)   - \frac{1}{8\pi\rho}\frac{\de B^2}{\de R},\\
&& \frac{\de \vphi}{\de t} + \vv\cdot\nabla\vphi + \frac{\vR \vphi}{R}= -\frac{1}{\rho R}\frac{\de p}{\de \phi} 
+ \frac{1}{4\pi\rho}\left( \Bv\cdot\nabla\Bphi + \frac{\BR \Bphi}{R}\right)  - \frac{1}{8\pi\rho}\frac{1}{R}\frac{\de B^2}{\de \phi}, \\
&& \frac{\de \vz}{\de t} + \vv\cdot\nabla\vz = -\frac{1}{\rho}\frac{\de p}{\de z} - \frac{\de \Phi}{\de z}  
+ \frac{1}{4\pi\rho}\left( \Bv\cdot\nabla\Bz\right)  - \frac{1}{8\pi\rho}\frac{\de B^2}{\de z},\\
\label{eq:indR}
&& \frac{\de \BR}{\de t} = \Bv\cdot\nabla\vR - \vv\cdot\nabla\BR  - (\div \vv)\BR,\\
\label{eq:indphi}
&& \frac{\de \Bphi}{\de t} =  \Bv\cdot\nabla\vphi + \frac{\Bphi \vR}{R} -  \vv\cdot\nabla\Bphi 
- \frac{\vphi \BR}{R} - (\div \vv)\Bphi,\\
\label{eq:indz}
&& \frac{\de \Bz}{\de t} =  \Bv\cdot\nabla\vz -  \vv\cdot\nabla\Bz - (\div \vv)\Bz, \\
&& \frac{p}{\gamma -1}\left(\frac{\de}{\de t} + \vv\cdot\nabla\right) \ln (p\rho^{-\gamma}) = \frac{1}{R}\frac{\de}{\de R}\left[R\frac{\chi(T) \BR (\Bv\cdot\nabla)T}{B^2} \right]
+ \frac{1}{R}\frac{\de}{\de \phi}\left[\frac{\chi(T) \Bphi (\Bv\cdot\nabla)T}{B^2} \right]
\nonumber \\ && \qquad + \frac{\de}{\de z} \left[\frac{\chi(T) \Bz (\Bv\cdot\nabla)T}{B^2}\right] 
- \rho\L(T,\rho) +\H(R,z),
\label{eq:model_cyl2}
\end{eqnarray}
where we have assumed that the gravitational potential is axisymmetric
and we have used $\div \Bv = 0$ in writing the three components
(equations ~\ref{eq:indR}-\ref{eq:indz}) of the induction equation
(\ref{eq:indu}). 

\section{Hydrostatic mass bias}\label{app:bias}

The estimate of the mass of galaxy clusters is mainly based on X-ray
and gravitational lensing data.  Mass estimates from X-rays are
obtained in general under the assumption that the gas is in
hydrostatic equilibrium, while mass estimates from gravitational
lensing do not depend on the kinematics of gas.  Deviations from
hydrostatic equilibrium (due to rotation, bulk motions or turbulence)
produce a discrepancy between the mass estimated from X-ray data and
the mass estimated from gravitational lensing. Such a discrepancy,
which is usually called hydrostatic mass bias, can be quantified by
measuring within a sphere of radius $r$ the ratio
$\Mtherm(r)/\Mtrue(r)$, where $\Mtrue$ is the true mass and $\Mtherm$
is the mass computed assuming that gravity is balanced only by the
thermal-pressure gradient \citep{Lau13}.

The ratio $\Mtherm/\Mtrue$ is estimated in hydrodynamic cosmological
simulations to be around $0.8-0.9$, with a slight variation with
radius out to $r_{200}$ and a scatter of about 20 per cent
\citep{Men10,Nel14}. When X-ray simulated observations are used to
mimic typical X-ray exposures, \citet{Ras12} show that temperature
inhomogeneities could affect further the hydrostatic mass
reconstruction by reducing $\Mtherm/\Mtrue$ to $0.65-0.75$.  On the
observational side, we cannot probe the ``true'' mass but only a
reasonable proxy for it as provided by gravitational lensing
estimates. Programs like Weighing the Giants (WtG, \citealt{von14}),
CLASH \citep{Don14}, and the Canadian Cluster Comparison Project
(CCCP, \citealt{Hoe15}) obtain values of the hydrostatic-to-lensing
mass ratio in the range of $0.7-0.8$, although differences up to 40
per cent in either the weak lensing or X-ray mass measurement among
different research groups are still measured \citep{Ser15}.

\begin{figure}
  \centerline{\includegraphics{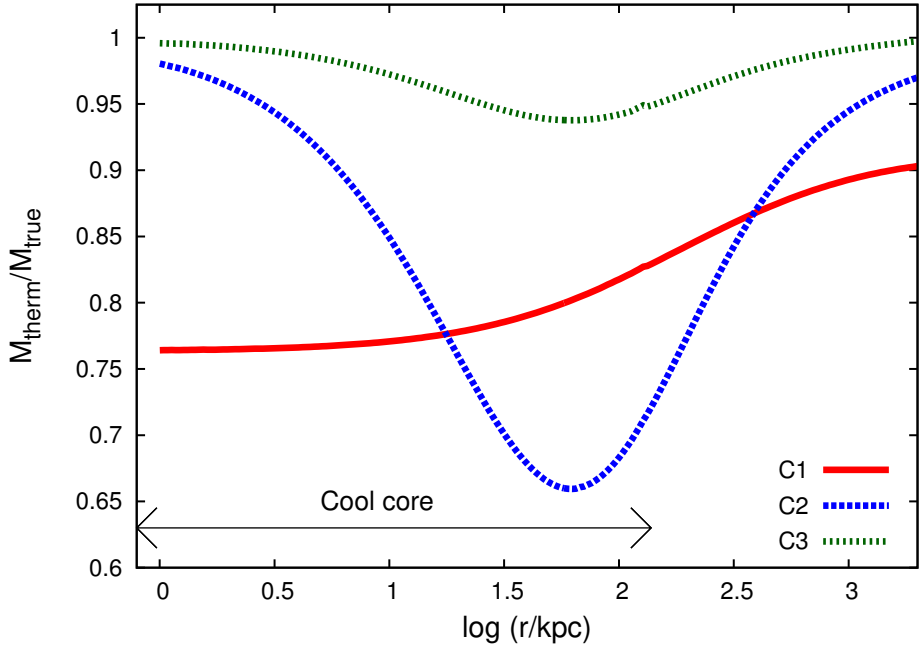}}
  \caption{Hydrostatic ($\Mtherm$) to true ($\Mtrue$) mass ratio
    within a sphere of radius $r$ for the same models as in
    Fig.~\ref{fig:nezero}.}
\label{fig:bias}
\end{figure}

For models of rotating cluster such as C1, C2 and C3 (see
Section~\ref{sec:unp}) $\Mtherm<\Mtrue$, because of the presence of
rotational support. Models in which rotational support is more
important are characterized by lower values of $\Mtherm/\Mtrue$: for a
rotating model to be realistic, it must have $\Mtherm/\Mtrue$ higher
than the aforementioned observed values. It is therefore interesting
to estimate the hydrostatic mass bias for models C1, C2 and C3. In all
three cases $\Mtrue(r)$ is just given by the spherically symmetric NFW
mass distribution responsible for the total cluster potential (see
Section~\ref{sec:unp}). Following \citet{Lau13}, the hydrostatic mass
profile can be computed as
\begin{equation}
 \Mtherm(r)\equiv -\frac{r^2}{G\av{\rhozero}}\frac{\de \av{\pzero}}{\de r},
\end{equation}
where the symbol $\av{\cdots}$ indicates an average over angles at
fixed spherical radial coordinate $r$.  We construct $\av{\rhozero}$
and $\av{\pzero}$ by interpolating $\rhozero(R,Z)$ and $\pzero(R,z)$
in spherical coordinates $(r,\theta)$ and then computing the average
over $\theta$, at fixed $r$. The profiles of $\Mtherm/\Mtrue$ for
models C1, C2 and C3 are shown in Fig.~\ref{fig:bias}.  The minimum
values of $\Mtherm/\Mtrue$ are 0.76, 0.66 and 0.94, for models C1, C2
and C3, respectively. However, these minima occur within the cool-core
region ($r<100\kpc$), in which the hydrostatic mass bias is not easily
measured observationally. The minimum values of $\Mtherm/\Mtrue$ in
the radial range $\rs\leq r \leq r_{200}$ (i.e. $0.5\lesssim r/\Mpc
\lesssim 2$), which is representative of the region typically probed
by weak gravitational lensing, are 0.88 (C1), 0.90 (C2) and 0.98 (C3),
consistent with the observational limits.

\bibliographystyle{jpp}


\bibliography{nipotietal_jpp_2015_final}

\begin{thebibliography}{51}
\expandafter\ifx\csname natexlab\endcsname\relax\def\natexlab#1{#1}\fi

\bibitem[{Balbus}(2000)]{Bal00}
{\sc {Balbus}, S.~A.} 2000 {Stability, Instability, and ``Backward'' Transport
  in Stratified Fluids}. {\em \apj\/} {\bf 534}, 420--427.

\bibitem[{Balbus}(2001)]{Bal01}
{\sc {Balbus}, S.~A.} 2001 {Convective and Rotational Stability of a Dilute
  Plasma}. {\em \apj\/} {\bf 562}, 909--917.

\bibitem[{Balbus}(2004)]{Bal04}
{\sc {Balbus}, S.~A.} 2004 {Viscous Shear Instability in Weakly Magnetized,
  Dilute Plasmas}. {\em \apj\/} {\bf 616}, 857--864.

\bibitem[{Balbus} {\em et~al.\/}(1994){Balbus}, {Gammie} \& {Hawley}]{Bal94}
{\sc {Balbus}, S.~A., {Gammie}, C.~F. \& {Hawley}, J.~F.} 1994 {Fluctuations,
  dissipation and turbulence in accretion discs}. {\em \mnras\/} {\bf 271},
  197.

\bibitem[{Balbus} \& {Hawley}(1991)]{BalH91}
{\sc {Balbus}, S.~A. \& {Hawley}, J.~F.} 1991 {A powerful local shear
  instability in weakly magnetized disks. I - Linear analysis. II - Nonlinear
  evolution}. {\em \apj\/} {\bf 376}, 214--233.

\bibitem[{Bianconi} {\em et~al.\/}(2013){Bianconi}, {Ettori} \&
  {Nipoti}]{Bia13}
{\sc {Bianconi}, M., {Ettori}, S. \& {Nipoti}, C.} 2013 {Gas rotation in galaxy
  clusters: signatures and detectability in X-rays}. {\em \mnras\/} {\bf 434},
  1565--1575 (BEN13).

\bibitem[{Binney} {\em et~al.\/}(2009){Binney}, {Nipoti} \&
  {Fraternali}]{Bin09}
{\sc {Binney}, J., {Nipoti}, C. \& {Fraternali}, F.} 2009 {Do high-velocity
  clouds form by thermal instability?} {\em \mnras\/} {\bf 397}, 1804--1815.

\bibitem[{Braginskii}(1965)]{Bra65}
{\sc {Braginskii}, S.~I.} 1965 {Transport Processes in a Plasma}. {\em Reviews
  of Plasma Physics\/} {\bf 1}, 205.

\bibitem[{Br{\"u}ggen}(2013)]{Bru13}
{\sc {Br{\"u}ggen}, M.} 2013 {Magnetic fields in galaxy clusters}. {\em
  Astronomische Nachrichten\/} {\bf 334}, 543.

\bibitem[{Chandrasekhar}(1960)]{Cha60}
{\sc {Chandrasekhar}, S.} 1960 {The Stability of Non-Dissipative Couette Flow
  in Hydromagnetics}. {\em Proceedings of the National Academy of Science\/}
  {\bf 46}, 253--257.

\bibitem[{Ciotti} \& {Ostriker}(2007)]{Cio07}
{\sc {Ciotti}, L. \& {Ostriker}, J.~P.} 2007 {Radiative Feedback from Massive
  Black Holes in Elliptical Galaxies: AGN Flaring and Central Starburst Fueled
  by Recycled Gas}. {\em \apj\/} {\bf 665}, 1038--1056.

\bibitem[{Donahue} {\em et~al.\/}(2014){Donahue}, {Voit}, {Mahdavi}, {Umetsu},
  {Ettori}, {Merten}, {Postman}, {Hoffer}, {Baldi}, {Coe}, {Czakon},
  {Bartelmann}, {Benitez}, {Bouwens}, {Bradley}, {Broadhurst}, {Ford},
  {Gastaldello}, {Grillo}, {Infante}, {Jouvel}, {Koekemoer}, {Kelson}, {Lahav},
  {Lemze}, {Medezinski}, {Melchior}, {Meneghetti}, {Molino}, {Moustakas},
  {Moustakas}, {Nonino}, {Rosati}, {Sayers}, {Seitz}, {Van der Wel}, {Zheng} \&
  {Zitrin}]{Don14}
{\sc {Donahue}, M., {Voit}, G.~M., {Mahdavi}, A., {Umetsu}, K., {Ettori}, S.,
  {Merten}, J., {Postman}, M., {Hoffer}, A., {Baldi}, A., {Coe}, D., {Czakon},
  N., {Bartelmann}, M., {Benitez}, N., {Bouwens}, R., {Bradley}, L.,
  {Broadhurst}, T., {Ford}, H., {Gastaldello}, F., {Grillo}, C., {Infante}, L.,
  {Jouvel}, S., {Koekemoer}, A., {Kelson}, D., {Lahav}, O., {Lemze}, D.,
  {Medezinski}, E., {Melchior}, P., {Meneghetti}, M., {Molino}, A.,
  {Moustakas}, J., {Moustakas}, L.~A., {Nonino}, M., {Rosati}, P., {Sayers},
  J., {Seitz}, S., {Van der Wel}, A., {Zheng}, W. \& {Zitrin}, A.} 2014
  {CLASH-X: A Comparison of Lensing and X-Ray Techniques for Measuring the Mass
  Profiles of Galaxy Clusters}. {\em \apj\/} {\bf 794}, 136.

\bibitem[{Ettori} {\em et~al.\/}(2013){Ettori}, {Pratt}, {de Plaa}, {Eckert},
  {Nevalainen}, {Battistelli}, {Borgani}, {Croston}, {Finoguenov}, {Kaastra},
  {Gaspari}, {Gastaldello}, {Gitti}, {Molendi}, {Pointecouteau}, {Ponman},
  {Reiprich}, {Roncarelli}, {Rossetti}, {Sanders}, {Sun}, {Trinchieri},
  {Vazza}, {Arnaud}, {B{\"o}ringher}, {Brighenti}, {Dahle}, {De Grandi},
  {Mohr}, {Moretti} \& {Schindler}]{Ett13}
{\sc {Ettori}, S., {Pratt}, G.~W., {de Plaa}, J., {Eckert}, D., {Nevalainen},
  J., {Battistelli}, E.~S., {Borgani}, S., {Croston}, J.~H., {Finoguenov}, A.,
  {Kaastra}, J., {Gaspari}, M., {Gastaldello}, F., {Gitti}, M., {Molendi}, S.,
  {Pointecouteau}, E., {Ponman}, T.~J., {Reiprich}, T.~H., {Roncarelli}, M.,
  {Rossetti}, M., {Sanders}, J.~S., {Sun}, M., {Trinchieri}, G., {Vazza}, F.,
  {Arnaud}, M., {B{\"o}ringher}, H., {Brighenti}, F., {Dahle}, H., {De Grandi},
  S., {Mohr}, J.~J., {Moretti}, A. \& {Schindler}, S.} 2013 {The Hot and
  Energetic Universe: The astrophysics of galaxy groups and clusters}. {\em
  ArXiv e-prints, arXiv:1306.2322\/} .

\bibitem[{Fang} {\em et~al.\/}(2009){Fang}, {Humphrey} \& {Buote}]{Fan09}
{\sc {Fang}, T., {Humphrey}, P. \& {Buote}, D.} 2009 {Rotation and Turbulence
  of the Hot Intracluster Medium in Galaxy Clusters}. {\em \apj\/} {\bf 691},
  1648--1659.

\bibitem[{Ferraro}(1937)]{Fer37}
{\sc {Ferraro}, V.~C.~A.} 1937 {The non-uniform rotation of the Sun and its
  magnetic field}. {\em \mnras\/} {\bf 97}, 458.

\bibitem[{Field}(1965)]{Fie65}
{\sc {Field}, G.~B.} 1965 {Thermal Instability.} {\em \apj\/} {\bf 142}, 531.

\bibitem[{Gaspari} {\em et~al.\/}(2012){Gaspari}, {Ruszkowski} \&
  {Sharma}]{Gas12}
{\sc {Gaspari}, M., {Ruszkowski}, M. \& {Sharma}, P.} 2012 {Cause and Effect of
  Feedback: Multiphase Gas in Cluster Cores Heated by AGN Jets}. {\em \apj\/}
  {\bf 746}, 94.

\bibitem[{Heckman}(1981)]{Hec81}
{\sc {Heckman}, T.~M.} 1981 {Optical emission-line gas associated with dominant
  cluster galaxies}. {\em \apjl\/} {\bf 250}, L59--L63.

\bibitem[{Hoekstra} {\em et~al.\/}(2015){Hoekstra}, {Herbonnet}, {Muzzin},
  {Babul}, {Mahdavi}, {Viola} \& {Cacciato}]{Hoe15}
{\sc {Hoekstra}, H., {Herbonnet}, R., {Muzzin}, A., {Babul}, A., {Mahdavi}, A.,
  {Viola}, M. \& {Cacciato}, M.} 2015 {The Canadian Cluster Comparison Project:
  detailed study of systematics and updated weak lensing masses}. {\em
  \mnras\/} {\bf 449}, 685--714.

\bibitem[{Kitayama} {\em et~al.\/}(2014){Kitayama}, {Bautz}, {Markevitch},
  {Matsushita}, {Allen}, {Kawaharada}, {McNamara}, {Ota}, {Akamatsu}, {de
  Plaa}, {Galeazzi}, {Madejski}, {Main}, {Miller}, {Nakazawa}, {Russell},
  {Sato}, {Sekiya}, {Simionescu}, {Tamura}, {Uchida}, {Ursino}, {Werner},
  {Zhuravleva}, {ZuHone} \& {on behalf of the ASTRO-H Science Working
  Group}]{Kit14}
{\sc {Kitayama}, T., {Bautz}, M., {Markevitch}, M., {Matsushita}, K., {Allen},
  S., {Kawaharada}, M., {McNamara}, B., {Ota}, N., {Akamatsu}, H., {de Plaa},
  J., {Galeazzi}, M., {Madejski}, G., {Main}, R., {Miller}, E., {Nakazawa}, K.,
  {Russell}, H., {Sato}, K., {Sekiya}, N., {Simionescu}, A., {Tamura}, T.,
  {Uchida}, Y., {Ursino}, E., {Werner}, N., {Zhuravleva}, I., {ZuHone}, J. \&
  {on behalf of the ASTRO-H Science Working Group}} 2014 {ASTRO-H White Paper -
  Clusters of Galaxies and Related Science}. {\em ArXiv e-prints,
  arXiv:1412.1176\/} .

\bibitem[{Kunz}(2011)]{Kun11}
{\sc {Kunz}, M.~W.} 2011 {Dynamical stability of a thermally stratified
  intracluster medium with anisotropic momentum and heat transport}. {\em
  \mnras\/} {\bf 417}, 602--616.

\bibitem[{Kunz} {\em et~al.\/}(2012){Kunz}, {Bogdanovi{\'c}}, {Reynolds} \&
  {Stone}]{Kun12}
{\sc {Kunz}, M.~W., {Bogdanovi{\'c}}, T., {Reynolds}, C.~S. \& {Stone}, J.~M.}
  2012 {Buoyancy Instabilities in a Weakly Collisional Intracluster Medium}.
  {\em \apj\/} {\bf 754}, 122.

\bibitem[{Latter} \& {Kunz}(2012)]{Lat12}
{\sc {Latter}, H.~N. \& {Kunz}, M.~W.} 2012 {The HBI in a quasi-global model of
  the intracluster medium}. {\em \mnras\/} {\bf 423}, 1964--1972.

\bibitem[{Lau} {\em et~al.\/}(2012){Lau}, {Nagai}, {Kravtsov}, {Vikhlinin} \&
  {Zentner}]{Lau12}
{\sc {Lau}, E.~T., {Nagai}, D., {Kravtsov}, A.~V., {Vikhlinin}, A. \&
  {Zentner}, A.~R.} 2012 {Constraining Cluster Physics with the Shape of X-Ray
  Clusters: Comparison of Local X-Ray Clusters Versus {$\Lambda$}CDM Clusters}.
  {\em \apj\/} {\bf 755}, 116.

\bibitem[{Lau} {\em et~al.\/}(2013){Lau}, {Nagai} \& {Nelson}]{Lau13}
{\sc {Lau}, E.~T., {Nagai}, D. \& {Nelson}, K.} 2013 {Weighing Galaxy Clusters
  with Gas. I. On the Methods of Computing Hydrostatic Mass Bias}. {\em \apj\/}
  {\bf 777}, 151.

\bibitem[{Malagoli} {\em et~al.\/}(1987){Malagoli}, {Rosner} \& {Bodo}]{Mal87}
{\sc {Malagoli}, A., {Rosner}, R. \& {Bodo}, G.} 1987 {On the thermal
  instability of galactic and cluster halos}. {\em \apj\/} {\bf 319}, 632--636.

\bibitem[{McCourt} {\em et~al.\/}(2011){McCourt}, {Parrish}, {Sharma} \&
  {Quataert}]{Mcc11}
{\sc {McCourt}, M., {Parrish}, I.~J., {Sharma}, P. \& {Quataert}, E.} 2011 {Can
  conduction induce convection? On the non-linear saturation of buoyancy
  instabilities in dilute plasmas}. {\em \mnras\/} {\bf 413}, 1295--1310.

\bibitem[{McCourt} {\em et~al.\/}(2012){McCourt}, {Sharma}, {Quataert} \&
  {Parrish}]{Mcc12}
{\sc {McCourt}, M., {Sharma}, P., {Quataert}, E. \& {Parrish}, I.~J.} 2012
  {Thermal instability in gravitationally stratified plasmas: implications for
  multiphase structure in clusters and galaxy haloes}. {\em \mnras\/} {\bf
  419}, 3319--3337.

\bibitem[{Meneghetti} {\em et~al.\/}(2010){Meneghetti}, {Rasia}, {Merten},
  {Bellagamba}, {Ettori}, {Mazzotta}, {Dolag} \& {Marri}]{Men10}
{\sc {Meneghetti}, M., {Rasia}, E., {Merten}, J., {Bellagamba}, F., {Ettori},
  S., {Mazzotta}, P., {Dolag}, K. \& {Marri}, S.} 2010 {Weighing simulated
  galaxy clusters using lensing and X-ray}. {\em \aap\/} {\bf 514}, A93.

\bibitem[{Nagai} {\em et~al.\/}(2013){Nagai}, {Lau}, {Avestruz}, {Nelson} \&
  {Rudd}]{Nag13}
{\sc {Nagai}, D., {Lau}, E.~T., {Avestruz}, C., {Nelson}, K. \& {Rudd}, D.~H.}
  2013 {Predicting Merger-induced Gas Motions in {$\Lambda$}CDM Galaxy
  Clusters}. {\em \apj\/} {\bf 777}, 137.

\bibitem[{Navarro} {\em et~al.\/}(1995){Navarro}, {Frenk} \& {White}]{Nav95}
{\sc {Navarro}, J.~F., {Frenk}, C.~S. \& {White}, S.~D.~M.} 1995 {Simulations
  of X-ray clusters}. {\em \mnras\/} {\bf 275}, 720--740.

\bibitem[{Nelson} {\em et~al.\/}(2014){Nelson}, {Lau}, {Nagai}, {Rudd} \&
  {Yu}]{Nel14}
{\sc {Nelson}, K., {Lau}, E.~T., {Nagai}, D., {Rudd}, D.~H. \& {Yu}, L.} 2014
  {Weighing Galaxy Clusters with Gas. II. On the Origin of Hydrostatic Mass
  Bias in {$\Lambda$}CDM Galaxy Clusters}. {\em \apj\/} {\bf 782}, 107.

\bibitem[{Nipoti}(2010)]{Nip10}
{\sc {Nipoti}, C.} 2010 {Thermal instability in rotating galactic coronae}.
  {\em \mnras\/} {\bf 406}, 247--263.

\bibitem[{Nipoti} \& {Binney}(2004)]{Nip04}
{\sc {Nipoti}, C. \& {Binney}, J.} 2004 {Cold filaments in galaxy clusters:
  effects of heat conduction}. {\em \mnras\/} {\bf 349}, 1509--1515.

\bibitem[{Nipoti} \& {Posti}(2013)]{Nip13}
{\sc {Nipoti}, C. \& {Posti}, L.} 2013 {Thermal stability of a weakly
  magnetized rotating plasma}. {\em \mnras\/} {\bf 428}, 815--827 (NP13).

\bibitem[{Nipoti} \& {Posti}(2014)]{Nip14}
{\sc {Nipoti}, C. \& {Posti}, L.} 2014 {On the Nature of Local Instabilities in
  Rotating Galactic Coronae and Cool Cores of Galaxy Clusters}. {\em \apj\/}
  {\bf 792}, 21 (NP14).

\bibitem[{Parrish} {\em et~al.\/}(2012){Parrish}, {McCourt}, {Quataert} \&
  {Sharma}]{Par12}
{\sc {Parrish}, I.~J., {McCourt}, M., {Quataert}, E. \& {Sharma}, P.} 2012 {The
  effects of anisotropic viscosity on turbulence and heat transport in the
  intracluster medium}. {\em \mnras\/} {\bf 422}, 704--718.

\bibitem[{Parrish} {\em et~al.\/}(2009){Parrish}, {Quataert} \&
  {Sharma}]{Par09}
{\sc {Parrish}, I.~J., {Quataert}, E. \& {Sharma}, P.} 2009 {Anisotropic
  Thermal Conduction and the Cooling Flow Problem in Galaxy Clusters}. {\em
  \apj\/} {\bf 703}, 96--108.

\bibitem[{Pinto} {\em et~al.\/}(2015){Pinto}, {Sanders}, {Werner}, {de Plaa},
  {Fabian}, {Zhang}, {Kaastra}, {Finoguenov} \& {Ahoranta}]{Pin15}
{\sc {Pinto}, C., {Sanders}, J.~S., {Werner}, N., {de Plaa}, J., {Fabian},
  A.~C., {Zhang}, Y.-Y., {Kaastra}, J.~S., {Finoguenov}, A. \& {Ahoranta}, J.}
  2015 {Chemical Enrichment RGS cluster Sample (CHEERS): Constraints on
  turbulence}. {\em \aap\/} {\bf 575}, A38.

\bibitem[{Quataert}(2008)]{Qua08}
{\sc {Quataert}, E.} 2008 {Buoyancy Instabilities in Weakly Magnetized
  Low-Collisionality Plasmas}. {\em \apj\/} {\bf 673}, 758--762.

\bibitem[{Quataert} {\em et~al.\/}(2002){Quataert}, {Dorland} \&
  {Hammett}]{Qua02}
{\sc {Quataert}, E., {Dorland}, W. \& {Hammett}, G.~W.} 2002 {The
  Magnetorotational Instability in a Collisionless Plasma}. {\em \apj\/} {\bf
  577}, 524--533.

\bibitem[{Rasia} {\em et~al.\/}(2012){Rasia}, {Meneghetti}, {Martino},
  {Borgani}, {Bonafede}, {Dolag}, {Ettori}, {Fabjan}, {Giocoli}, {Mazzotta},
  {Merten}, {Radovich} \& {Tornatore}]{Ras12}
{\sc {Rasia}, E., {Meneghetti}, M., {Martino}, R., {Borgani}, S., {Bonafede},
  A., {Dolag}, K., {Ettori}, S., {Fabjan}, D., {Giocoli}, C., {Mazzotta}, P.,
  {Merten}, J., {Radovich}, M. \& {Tornatore}, L.} 2012 {Lensing and x-ray mass
  estimates of clusters (simulations)}. {\em New Journal of Physics\/} {\bf
  14}~(5), 055018.

\bibitem[{Sanders} \& {Fabian}(2013)]{San13}
{\sc {Sanders}, J.~S. \& {Fabian}, A.~C.} 2013 {Velocity width measurements of
  the coolest X-ray emitting material in the cores of clusters, groups and
  elliptical galaxies}. {\em \mnras\/} {\bf 429}, 2727--2738.

\bibitem[{Sereno} \& {Ettori}(2015)]{Ser15}
{\sc {Sereno}, M. \& {Ettori}, S.} 2015 {Comparing Masses in Literature
  (CoMaLit)-I. Bias and scatter in weak lensing and X-ray mass estimates of
  clusters}. {\em \mnras, in press (arXiv:1407.7868v2)\/} .

\bibitem[{Spitzer}(1962)]{Spi62}
{\sc {Spitzer}, L.} 1962 {\em {Physics of Fully Ionized Gases}\/}. New York:
  Interscience (2nd edition).

\bibitem[{Sutherland} \& {Dopita}(1993)]{Sut93}
{\sc {Sutherland}, R.~S. \& {Dopita}, M.~A.} 1993 {Cooling functions for
  low-density astrophysical plasmas}. {\em \apjs\/} {\bf 88}, 253--327.

\bibitem[{Takahashi} {\em et~al.\/}(2014){Takahashi}, {Mitsuda}, {Kelley},
  {Fabian}, {Mushotzky}, {Ohashi}, {Petre} \& {on behalf of the ASTRO-H Science
  Working Group}]{Tak14}
{\sc {Takahashi}, T., {Mitsuda}, K., {Kelley}, R., {Fabian}, A., {Mushotzky},
  R., {Ohashi}, T., {Petre}, R. \& {on behalf of the ASTRO-H Science Working
  Group}} 2014 {ASTRO-H White Paper - Introduction}. {\em ArXiv e-prints,
  arXiv:1412.2351\/} .

\bibitem[{Velikhov}(1959)]{Vel59}
{\sc {Velikhov}, E.~P.} 1959 {Stability of an ideally conducting liquid flowing
  between cylinders rotating in a magnetic field}. {\em Soviet Phys. JETP\/}
  {\bf 9}, 995.

\bibitem[{Voit} \& {Donahue}(2015)]{Voi15}
{\sc {Voit}, G.~M. \& {Donahue}, M.} 2015 {Cooling Time, Freefall Time, and
  Precipitation in the Cores of ACCEPT Galaxy Clusters}. {\em \apjl\/} {\bf
  799}, L1.

\bibitem[{von der Linden} {\em et~al.\/}(2014){von der Linden}, {Mantz},
  {Allen}, {Applegate}, {Kelly}, {Morris}, {Wright}, {Allen}, {Burchat},
  {Burke}, {Donovan} \& {Ebeling}]{von14}
{\sc {von der Linden}, A., {Mantz}, A., {Allen}, S.~W., {Applegate}, D.~E.,
  {Kelly}, P.~L., {Morris}, R.~G., {Wright}, A., {Allen}, M.~T., {Burchat},
  P.~R., {Burke}, D.~L., {Donovan}, D. \& {Ebeling}, H.} 2014 {Robust
  weak-lensing mass calibration of Planck galaxy clusters}. {\em \mnras\/} {\bf
  443}, 1973--1978.

\bibitem[{Zhuravleva} {\em et~al.\/}(2014){Zhuravleva}, {Churazov},
  {Schekochihin}, {Allen}, {Ar{\'e}valo}, {Fabian}, {Forman}, {Sanders},
  {Simionescu}, {Sunyaev}, {Vikhlinin} \& {Werner}]{Zhu14}
{\sc {Zhuravleva}, I., {Churazov}, E., {Schekochihin}, A.~A., {Allen}, S.~W.,
  {Ar{\'e}valo}, P., {Fabian}, A.~C., {Forman}, W.~R., {Sanders}, J.~S.,
  {Simionescu}, A., {Sunyaev}, R., {Vikhlinin}, A. \& {Werner}, N.} 2014
  {Turbulent heating in galaxy clusters brightest in X-rays}. {\em \nat\/} {\bf
  515}, 85--87.

\end{thebibliography}

\end{document}